\documentclass[prl,twocolumn,floatfix,showpacs,superscriptaddress,preprintnumbers]{revtex4-1}
\usepackage{graphicx}
\usepackage[latin2]{inputenc}
\usepackage{amsmath}
\usepackage{amssymb}
\usepackage{epsfig}
\usepackage{bm}
\usepackage{color}

\begin{document}
\title{Topotaxis of Active Particles Induced by Spatially Heterogeneous Sliding along Obstacles}

\author{Zeinab Sadjadi}
\email{sadjadi@lusi.uni-sb.de}
\affiliation{Department of Theoretical Physics $\&$ Center for Biophysics, 
Saarland University, 66123 Saarbr\"ucken, Germany}
\author{Heiko Rieger}
\affiliation{Department of Theoretical Physics $\&$ Center for Biophysics, 
Saarland University, 66123 Saarbr\"ucken, Germany}
\affiliation{Leibniz Institute for New Materials INM, 66123 Saarbr\"ucken, Germany}

\begin{abstract}
Many biological active agents respond to gradients of environmental cues 
by redirecting their motion. Besides the well-studied prominent examples 
such as photo- and chemotaxis, there has been considerable recent interest 
in topotaxis, i.e.\ the ability to sense and follow topographic environmental 
cues. We numerically investigate the topotaxis of active agents moving in 
regular arrays of circular pillars. While a trivial topotaxis is achievable 
through a spatial gradient of obstacle density, here we show that imposing 
a gradient in the characteristics of agent-obstacle interaction can lead to 
an effective topotaxis in an environment with a spatially uniform density of 
obstacles. As a proof of concept, we demonstrate how a gradient in the angle 
of sliding around pillars--- as e.g.\ observed in bacterial dynamics near 
surfaces--- breaks the spatial symmetry and biases the direction of motion. 
We provide an explanation for this phenomenon based on effective reflection 
at the imaginary interface between pillars with different sliding angles. 
Our results are of technological importance for design of efficient taxis 
devices. 
\end{abstract}

\maketitle
Biological microswimmers, migrating cells, and other living organisms 
can sense environmental cues and external fields and respond by adapting 
their dynamics. The redirected motion in response 
to a gradient of a stimulus, called taxis, is a vital navigation 
mechanism in many biological systems. Topotaxis--- the ability to 
sense and follow topographic cues--- has attracted considerable 
attention \cite{Park18,Wondergem21,Reversat20,Gorelashvili14,
Schakenraad20,Frey06,Kaiser06,Muthinja17}, as it does not rely on 
the influence of any specific stimulus on the internal self-propulsion 
mechanism of the agent; it is solely based on the physical interactions 
with and properties of the surrounding environment such as spatial 
arrangement of obstacles, degree of lateral confinement, and surface 
topography. For a more efficient navigation, these features may be 
exploited by biological organisms--- particularly by immune cells, 
as they are responsible to explore extracellular matrices and confined 
tissues to detect pathogens \cite{Chabaud15,Shaebani20,Maiuri15,
Shaebani22a,Stankevicins20,Shaebani22b}. So far, topotaxis has been 
reported in the presence of spatial gradient of either obstacle density 
\cite{Wondergem21,Park18,Reversat20,Gorelashvili14,Schakenraad20} 
or substrate topography (for free motion on surfaces) \cite{Frey06,
Kaiser06}. It is unclear whether spatial variation of other topological 
features, such as a diversity in the size of obstacles, can lead 
to an efficient taxis.

Living organisms interact with obstacles in different ways. For 
instance, swimming bacteria may be hydrodynamically captured by 
and slide along surfaces \cite{Spagnolie15,Jakuszeit19,Sipos15}, 
migrating or killer cells are often temporarily trapped near 
obstacles \cite{Wondergem21,Sadjadi22,Arcizet12,Zhou17}, and 
microalgae push their flagella against surfaces and scatter 
\cite{Kantsler13,Lushi17,Contino15}. While the diffusivity may 
be enhanced by sliding around the objects \cite{Jakuszeit19}, 
it is reduced by being trapped near obstacles or scattered 
from them \cite{Saxton87,Saxton94,Machta83,Sadjadi08,Tierno16,
Dagdug12}. A detailed understanding of how the existence or 
strength of topotaxis depends on the interplay between topographic 
cues and the nature of agent-obstacle interaction is currently 
lacking. 

Here we study the motion of active agents in obstacle parks 
consisting of regularly arranged circular pillars. The density 
of pillars is the same throughout the system to prevent possible 
drifts due to obstacle density variations. We impose a gradient 
of topographical stimulus by varying the particle-obstacle 
interactions throughout the obstacle park, which is implemented 
through the sliding angle around pillars. By performing extensive 
numerical simulations of a persistent random walk with two distinct 
states in the bulk and in the vicinity of obstacles \cite{Shaebani22c,
Hafner16,Shaebani19}, we verify that the interplay between self-propulsion 
of the moving agents, agent-obstacle interactions, and topographical 
cues in the environment determines the possibility and strength 
of an effective topotaxis along the imposed gradients.

\begin{figure} [b]
\centering
\includegraphics[width=0.43\textwidth]{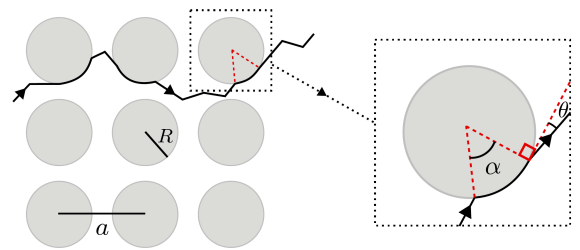}
\caption{Schematic drawing of a sample trajectory in a pillar array 
with lattice constant $a$ and pillar radius $R$. The sliding and 
leaving angles are denoted with $\alpha$ and $\theta$, respectively.}
\label{Fig1}
\end{figure} 
{\it Model.---} We consider a two-dimensional medium consisting of circular 
pillars with radius $R$. The pillars are placed on a square lattice with 
lattice constant $a$. We model the motion of the self-propelled agents by 
a persistent random walk. The walkers move with a constant step length 
$l$ at each time step. We define the mean local persistence as $p\,{=}\,
\langle\cos\phi\rangle$ \cite{Sadjadi21}, with $\phi$ being the turning 
angle between the successive steps and $\langle{\cdot}{\cdot}{\cdot}\rangle$ 
denoting the average with respect to the turning-angle distribution 
$R(\phi)$. The persistence values $p=1$ and $p=0$ correspond to ballistic 
and purely diffusive motion, respectively. We introduce a sliding boundary 
condition on the pillar surfaces. After collision, the walker moves along 
the obstacle surface with an angle $\alpha$ and leaves the obstacle surface 
with angle $\theta$ from tangent of the circle; see Fig.\,\ref{Fig1}. While 
the free parameter $\alpha$ can be pillar-size dependent in general 
\cite{Spagnolie15}, here we consider $\alpha$ to be independent $R$.

We perform Monte Carlo simulations of migration through the pillar park. 
The simulation box is $300l{\times}300l$, in which the pillars are arranged 
on a square lattice with lattice constant $a\,{=}\,12.5l$. A circular pillar 
with radius $R$ is placed on each lattice point (24 pillars in each row 
or column). By changing $R$, we vary the occupied fraction by pillars 
(characterized by the dimensionless parameter $\lambda\,{=}\,2R{/}a\,{\in}
\,[0,1]$). An event-driven algorithm is applied, where every collision with 
an obstacle is considered as a new event. The random walker takes a step 
with length $l$, unless it collides with an obstacle. In case of no collision, 
the walker takes a new direction drawn from the turning-angle distribution 
$R(\phi)$, which is chosen to be a uniform function over $[\phi_0,\phi_1]$. 
The values of the angles $\phi_0$ and $\phi_1$ can be tuned to get the 
desired persistence $p$. An ensemble of $10^5$ random walkers with random 
initial position and direction are considered and periodic boundary 
conditions are applied. To induce topotaxis, we consider a constant 
sliding angle $\alpha$ around each pillar but impose a gradient of 
$\alpha$ in the medium. 

\begin{figure} 
\centering
\includegraphics[scale=2.5,angle=0]{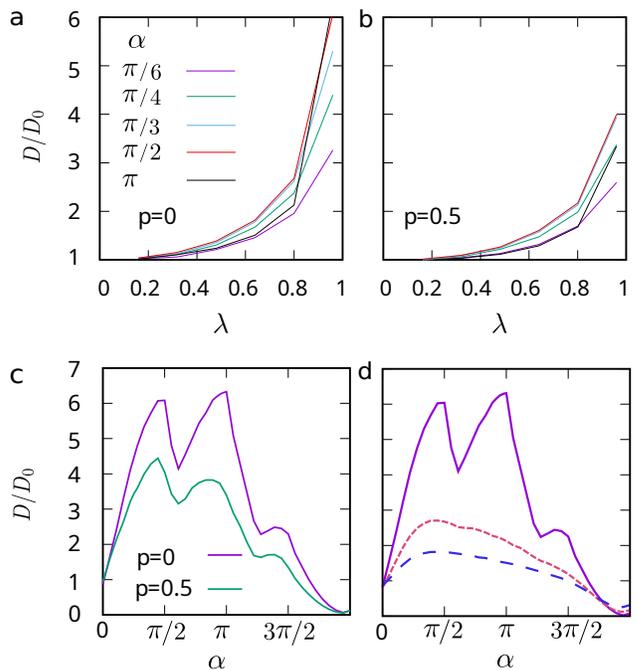}
\caption{(a,b) Rescaled diffusion constant $D{/}D_0$ 
versus $\lambda\,{=}\,2R/a$ for a (a) normal random walk 
and (b) persistent walk with persistence $p\,{=}\,0.5$. Each color represents 
a fixed sliding angle $\alpha$ on the obstacle surface. (c) Rescaled diffusion 
constant as a function of sliding angle $\alpha$ for a normal and persistent 
random walk. (d) $D{/}D_0$ vs $\alpha$ for normal random walk in pillar parks with 
various pillar densities. The full, dotted and dashed lines represent 
$\lambda\,{=}\,0.96, 0.8$ and $0.64$, respectively. In all cases the leaving 
angle $\theta$ is uniformly chosen from $[0,\frac{\pi}{4}]$.}
\label{Fig:F2}
\end{figure}

In order to understand the influence of the geometric parameters $\alpha$ and $\lambda$ on  
the particle migration in pillar parks, we study the behavior of the effective diffusion constant, $D$, at a given 
$\alpha$ and vary $\lambda$  by changing the radii of obstacles.
Since the diffusion constant in free space, $D_0$, depends on the persistence as $D_0{\propto}\frac{1+p}{1-p}$ 
\citep{Shaebani14}, we use the rescaled diffusion constant $\widetilde{D}{=}D{/}D_0$ in the following to eliminate the role of $p$. 
In Fig.\ref{Fig:F2}(a,b), $\widetilde{D}$ is plotted versus relative packing fraction 
$\lambda$ for different values of $\alpha$. The results are presented for 
diffusion ($p\,{=}\,0$) and persistent random walk with $p\,{=}\,0.5$. We 
observe that the diffusion constant grows with $\lambda$ and its variation 
is affected by the value of the sliding angle $\alpha$. Without sliding along 
obstacle surfaces, e.g.\ with reflective boundary condition on pillar surfaces, $\widetilde{D}$ 
decrease with pillar density, which is a known result \citep{Dagdug12}; 
however, when sliding along obstacle surface is allowed, $\widetilde{D}$ interestingly 
increases with density. For dense packing, the displacement $R\,\!\alpha$ 
on the perimeter of a pillar is larger than the pillar spacing. 
Moreover, in denser pillar parks, where random walkers are trapped between 
pillars, they use the sliding on pillar surface to escape the traps and 
thus propagate faster between pillars. In the case of persistent random 
walk [Fig.\ref{Fig:F2}(b)], with the same $\alpha$ as for $p{=}0$,
we observe a weaker increase in the relative diffusion constant. This is because 
active agents are less frequently trapped between pillars due to their active motion, 
thus, the relative impact of sliding on diffusion coefficient is less pronounced. In Fig.\ref{Fig:F2}(c), 
$\widetilde{D}$ of normal and persistent random walkers is plotted versus the sliding 
angle in a dense pillar park with $\lambda\,{=}\,0.96$. We observe three 
peaks at multiples of $\frac{\pi}{2}$. The maximum value of $\widetilde{D}$ is located 
either at $\alpha\,{=}\,\frac{\pi}{2}$ or $\alpha\,{=}\,\pi$, depending on 
the persistence of the random walker. However, this behavior disappears in 
smaller packing fractions, as shown for $p\,{=}\,0$ in Fig.\ref{Fig:F2}(d). 
A similar trend is observed for $p\,{>}\,0$. Thus, for sufficiently large $\lambda$, 
the impact of geometrical properties of the pillars (e.g. $\alpha$) on the diffusion constant are more pronounced.  
Based on these findings, we hypothesize that a gradient of sliding angle 
through the pillar park at large packing fractions can leads to topotaxis.
\begin{figure} [t]
\centering
\includegraphics[scale=0.83,angle=0]{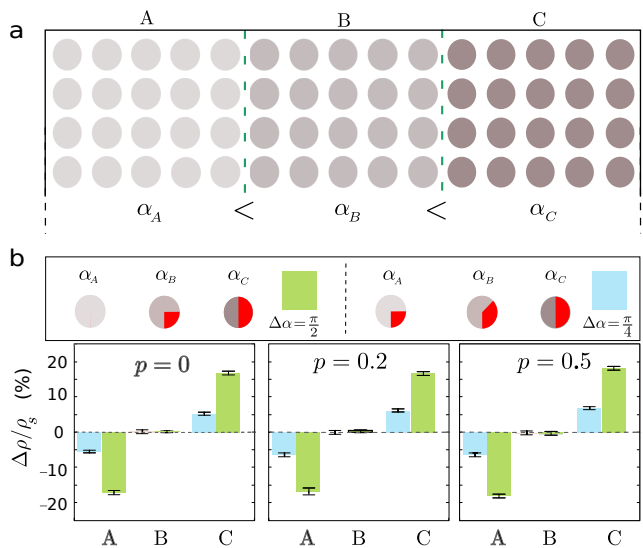}
\caption{ 
Inducing taxis in an obstacle park  with homogeneous packing fraction by varying 
the sliding angle. (a) Schematic representation of an obstacle park with gradient 
of sliding angle $\alpha$. A different value of $\alpha$ is assigned to each zone, 
with zone A (C) having the smallest (largest) angle. The color intensity of pillars 
is proportional to $\alpha$. (b) The relative particle density in each zone (with 
respect to the homogeneous stationary density $\rho_{_s}$) for normal and persistent 
random walks for two sets of $\alpha\!_{_A}, \alpha_{_B}$ and $\alpha_{_C}$ shown by 
green and blue colors. Here, $\lambda\,{=}\,0.96$ and $\theta$ is uniformly chosen 
from $[0,\frac{\pi}{4}]$. }
\label{Fig:taxis1}
\end{figure}

In order to induce topotaxis in a homogeneous medium with uniform packing fraction,
we consider a mono-disperse pillar park and assume that the sliding angle of particles 
along pillars is an angle which can be tuned by adapting the surface properties, 
e.g. by means of different coatings. We divide the pillar park into parallel sections 
and assign a constant sliding angle to each section. Here, we present the results for 
choosing three zones $A$, $B$, and $C$ with $\alpha\!_{_A}\,{<}\,\alpha_{_B}\,{<}\,\alpha_{_C}$, 
as depicted in Fig.\ref{Fig:taxis1}(a). Note that due to periodic boundary condition, 
sections A and C are also neighbors. We define particle density $\rho$ as the number of 
random walkers per unit of available area (i.e.\ pillar area excluded). We let the random 
walkers migrate in the pillar park, starting from random positions and orientations. 
In a homogeneous pillar park, one expects to have $\rho\!_{_A}\,{=}\,\rho_{_B}\,{=}\,\rho_{_C}\,{\equiv}\,\rho_{_s}$, 
with $\rho_{_s}{=}\,\frac{1}{3}$ being the particle density in the steady state of a 
homogeneous pillar park. Here, we observe $\rho\!_{_A}\,{<}\,\rho_{_B}\,{<}\,\rho_{_C}$, 
which means that the random walkers preferentially reside in the zones with larger 
sliding angles. Interestingly, this tendency mainly depends on the values of sliding angles 
(in a given $\lambda$) rather than the persistence of the random walker. A larger difference 
$\Delta\alpha$ between the sliding angles in adjacent zones results in a larger difference 
in the steady particle densities $\rho\!_{_A}$, $\rho_{_B}$, and $\rho_{_C}$. In Fig.\ref{Fig:taxis1}(b), 
exemplary variations of the relative density $\Delta\rho\,{=}\,\rho-\rho_{_s}$ are shown in different 
zones in the steady state. The results are presented for $p\,{=}\,0$, $0.2$ and $0.5$ and two  
sliding angles $\alpha\!_{_A}$, $\alpha_{_B}$ and $\alpha_{_C}$ as depicted by red color in the figure. 
For simplicity, we assume that the differences between sliding angles in the two neighboring zones are the 
same, i.e.\ $\Delta\alpha\,{=}\,\vert\alpha\!_{_A}{-}\,\alpha_{_B}\vert\,{=}\,\vert\alpha_{_B}{-}\,
\alpha_{_C}\vert\,{=}\,\frac{1}{2}\vert\alpha\!_{_A}{-}\,\alpha_{_C}\vert$. In examples shown in 
Fig.\ref{Fig:taxis1}(b) with green ad blue colors, $\Delta\alpha$ equals $\frac{\pi}{2}$ and 
$\frac{\pi}{4}$, respectively. For all choices of persistence $p$, a positive (negative) $\Delta\rho$ 
in section $C$ ($A$) in the steady state shows that more (less) particles are found there. Moreover, 
a larger $\Delta\alpha$ (green) results in a significantly larger $\vert \Delta\rho/\rho_{_s}\vert$, 
compared to the one with a smaller $\Delta\alpha$ (blue). This result indicates a taxis from smaller 
to larger $\alpha$ with a strength which depends on $\Delta\alpha$. We checked that other (inhomogeneous) 
initial conditions lead to similar conclusions.
The result illustrated by Fig.\ref{Fig:taxis1}(b) is somewhat counter-intuitive, since sliding with larger 
angles is reminiscent of active particles attaining higher activity or self-propulsion--- which 
leads in active Brownian particle systems to a depletion of particles \cite{Wysocki2022}, contrary to what happens here--- 
and we will clarify the reason below.

\begin{widetext}

\begin{figure*} [b]
\centering
\includegraphics[scale=3.3,angle=0]{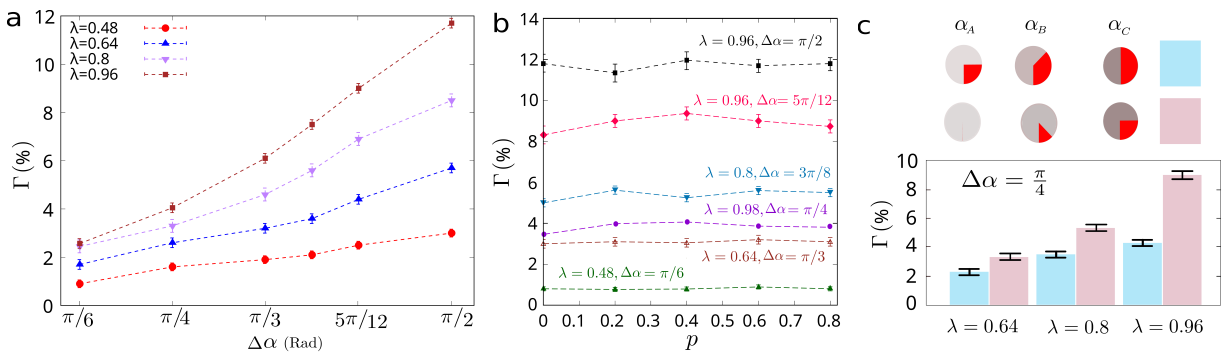}
\caption{(a) Topotaxis strength $\Gamma$ versus $\Delta\alpha$ for $p\,{=}\,0.6$ and different values of $\lambda$.
(b) $\Gamma$ vs persistence $p$ of random walker for various choices of $\lambda$ and $\Delta\alpha$.
(c) Topotaxis strength for $\Delta\alpha\,{=}\,\frac{\pi}{4}$ and two different choices of sliding angles. Here, $p\,{=}\,0.6$ and the results are shown for three
values of $\lambda$.  
}
\label{Fig:F4}
\end{figure*} 
\end{widetext}

To better understand how the topotaxis strength depends on $\Delta\alpha$ as well as the pillar packing fraction $\lambda$, 
we quantify the strength of taxis $\Gamma$ by the maximum difference between the steady densities, i.e.\  
$\Gamma\,{=}\,\rho_{_C}\,{-}\,\rho\!_{_A}$. In Fig.\ref{Fig:F4}(a), $\Gamma$
is plotted versus $\Delta\alpha$ for different values of $\lambda$ for a given $p$. We set $\alpha_{_C}{=}\,\pi$ and vary $\Delta\alpha$  
(i.e.\ choose $\alpha\!_{_A}\,{=}\,\pi\,{-}\,2\Delta\alpha$ and $\alpha\!_{_B}\,{=}\,\pi\,{-}\,\Delta\alpha$).
$\Gamma$ shows a nearly linear dependence on $\Delta\alpha$ which is stronger for larger $\lambda$. 
Even for middle values of $\lambda$, a significant topotaxis can be achieved by choosing 
proper parameters. Figure~\ref{Fig:F4}(b) 
shows $\Gamma$ versus $p$ for various choices of $\lambda$ and 
$\Delta\alpha$. It can be clearly seen that $\Gamma$ is independent of persistence, while 
it strongly depends on the choice of the geometrical parameters $\lambda$ 
and $\Delta\alpha$. We note that for a given choice 
of $\Delta\alpha$, there is a degree of freedom to choose the set of the sliding angles of the zones.
An example for $\Delta\alpha\,{=}\,\frac{\pi}{4}$ and
two choices of sliding angles is shown In Fig.\ref{Fig:F4}(c). 
Interestingly, $\Gamma$ depends not only on $\Delta\alpha$ but also on the chosen range of the sliding angles. 
Denoting the minimum sliding angle by $\alpha_{\text{min}}\,{\equiv}\,\alpha\!_{_A}$, we observe 
that choosing a smaller $\alpha_{\text{min}}$ leads to a larger $\Gamma$. Thus, for given values of $\Delta\alpha$ and $\lambda$, the 
maximum topotaxis strength is achieved for $\alpha_{\text{min}}\,{=}\,0$, i.e.\ no sliding on the pillars.

\begin{widetext}
 
 \begin{figure}[b]
\centering
\includegraphics[width=0.95\textwidth]{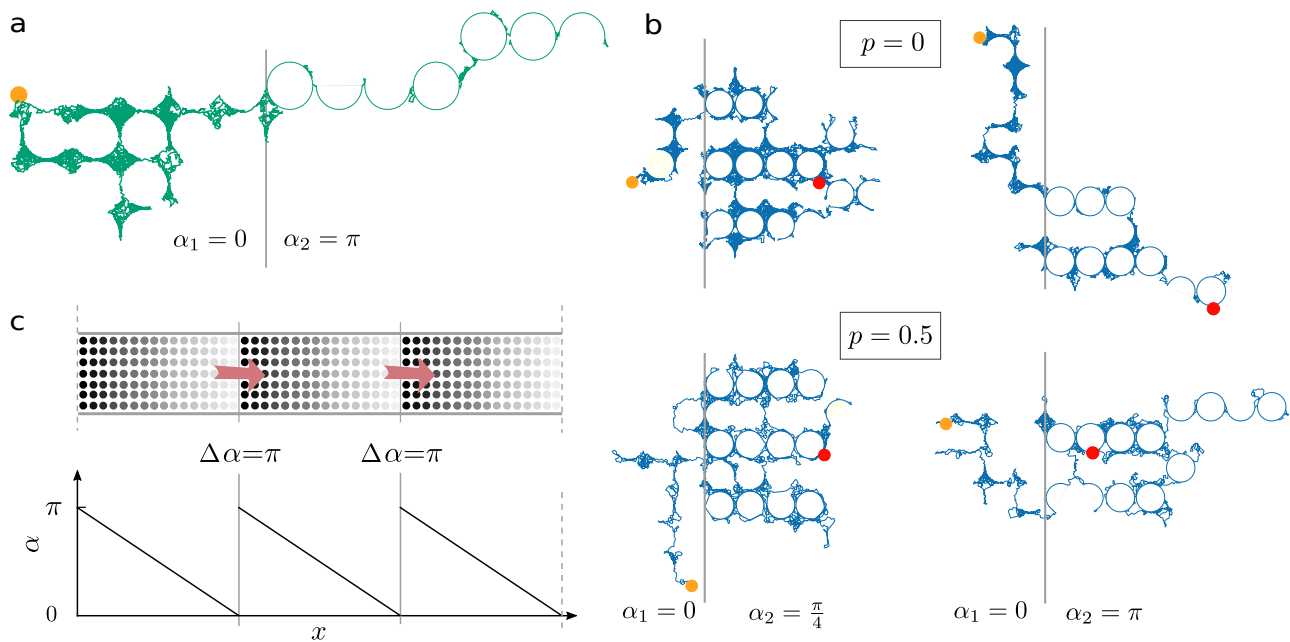}
\caption{(a) An exemplary trajectory at the interface of two zones with different $\alpha$. 
The starting point is depicted by the orange circle.
(b) Examples of longer trajectories where the random walker has enough time for several returns 
to the interface. Different panels represent either normal or persistent 
random walks for two choices of sliding angles $\alpha_1$ and $\alpha_2$. The orange and 
red circles represent the starting and final position of the random walker, respectively.
(c) Schematic design for guiding particles through blocks of pillars with a linear decrease of $\alpha$
in each block, represented with decreasing color intensity. Red arrows show the direction of the net flux. 
}
\label{Fig5}
\end{figure} 
\end{widetext}

In an inhomogeneous environment, random walkers tend to gather in regions where they have a lower mobility, 
i.e.\ smaller diffusion constant \cite{Schnitzer93}.
Therefore, the trivial way to induce topotaxis is to apply a gradient of packing fraction of obstacles, which changes
the local available space for migration. This way, the density of particles will be higher in regions with larger obstacle 
packing fraction, where particles have smaller $D$ due to frequent reflections. However, our findings demonstrate a counterintuitive possibility. 
We induce accumulation in zones with larger sliding angles, which have a larger diffusion constant [see Fig.\ref{Fig:F2}(c)]. 
To provide a qualitative understanding of the underlying mechanism, we focus on the interface between 
two zones with different sliding angles $\alpha_1\,{<}\,\alpha_2$. In Fig.\ref{Fig5}(a), a sample trajectory in the extreme case of $\alpha_1\,{=}\,0$ 
and $\alpha_2\,{=}\,\pi$ is depicted. Starting in region $1$ (i.e.\ left), the particle is often trapped between the obstacles due to frequent reflections from them. 
However, when it enters region 2 with the possibility of sliding on pillars, it can be effectively pulled 
into the medium by sliding around many pillar surfaces without being locally trapped.
If one waits enough, the particle returns to the interface again, as shown in Fig.\ref{Fig5}(b) for different choices of $p$, 
$\alpha_1$, and $\alpha_2$. While these sample random walkers have the chance to reenter region 1, the interface acts as a 
pseudo-reflective wall and effectively guides the walkers back to region 2.

Towards practical applications, such as guiding biological agents inside channels, particles should be guided in a 
specified direction over long distances. 
To this aim, one could partition the system into many blocks with successively increasing $\alpha$. However, this corresponds to 
small $\Delta\alpha$ between neighboring blocks, thus, a weak effective topotaxis strength $\Gamma$ [see Fig.\ref{Fig:F4}(a)];
indeed, $\Gamma{\rightarrow}0$ for $\Delta\alpha{\rightarrow}0$. We exploit this feature by linearly decreasing $\alpha$
from $\pi$ to $0$  in each block as shown in Fig.\ref{Fig5}(c). As a result, while the particles experience no significant topotaxis
within each block, they are pulled into the neighboring block  at the interface with $\Delta\,\alpha\,\,{=}\,\pi$. When

measures the net flux (by counting the number of particles crossing the interfaces between blocks in both directions)
in the simulation box, one obtains a significant flux of particles in the steady state, from left to right.

We note that the values of $\alpha_1$ and $\alpha_2$ and the trajectories in Fig.\ref{Fig5} 
have been selected to highlight the effective reflection at the interface. Nevertheless, transport 
occurs in both directions in general, though with asymmetric  probabilities $f\!_{_{1{\rightarrow}2}}$ and $f\!_{_{2{\rightarrow}1}}$
which depend on the geometrical parameters $\Delta\alpha$, $\alpha_{\text{min}}$, and $\lambda$ but are independent of $p$.
The Markov process of transport between these two zones eventually leads to steady state probabilities 
$\rho_{_1}{=}f\!_{_{2{\rightarrow}1}}/(f\!_{_{1{\rightarrow}2}}+f\!_{_{2{\rightarrow}1}})$ and $\rho_{_2}{=}f\!_{_{1{\rightarrow}2}}/(f\!_{_{1{\rightarrow}2}}+f\!_{_{2{\rightarrow}1}})$ for residence in each zone.
Although the explicit dependence of transition probabilities on topological properties of the medium 
is not known, their asymmetry is reflected in their ratio in the steady state, which is given as
$\frac{f\!_{_{1{\rightarrow}2}}}{f\!_{_{2{\rightarrow}1}}}{=}\frac{\rho_2}{\rho_1}$.
An strong topotaxis is gained for the set of conditions $\{\lambda{\rightarrow}1$, $\Delta\alpha{\gg}0$, and $\alpha_{\text{min}}{\rightarrow}0\}$.
On the other hand, in sparse pillar parks (i.e.\ $\lambda{\ll}1$) or in the limit of $\Delta\alpha{\rightarrow}0$,
we obtain $\rho_{_1}{\approx}\rho_{_2}{\approx}\rho_{_s}$ corresponding to a very weak topotaxis.
As a final note, while the persistence affects neither the transition probabilities nor the steady densities, it determines
the time scale to reach the steady state; a particle with a larger $p$ visits the interface more frequently
as it has a larger diffusion coefficient.

In summary, we have proposed a novel method to induce topotaxis in obstacle parks with a uniform packing fraction,
by applying a gradient in the angle of sliding around pillars. We verified that the method is capable of guiding 
agents over long distances. The concept can be generalized to other characteristics 
of agent-obstacle interactions or other geometrical properties of the environment. For example, we have preliminary results showing
that topotaxis can be also achieved in the absence of sliding by imposing a gradient of a degree of pillar-size polydispersity in the environment.
The persistence dependence of the relaxation time to the steady state can be exploited to separate a mixture of microorganisms with different persistence.
Our results are of technological importance as a non-invasive method (e.g. by imposing different pillar-surface coatings) to design taxis devices
for guiding biological agents.\\
 
\noindent This work was performed within and financially supported by the Collaborative Research Center SFB 1027  funded
by the German Research Foundation (DFG).\\
\bibliography{Refs}
\end{document}